\documentclass[runningheads]{llncs}

\usepackage[T1]{fontenc}
\usepackage{lmodern}
\usepackage{textcomp}

\usepackage{graphicx}
\usepackage{url}
\usepackage{subfig}

\usepackage{float}

\usepackage{booktabs}
\begin{document}

\title{Measuring football players' on-the-ball contributions from passes during games}
\titlerunning{Measuring football players' contributions from passes}

\author{Lotte Bransen \and Jan Van Haaren}
\authorrunning{L. Bransen and J. Van Haaren}

\institute{SciSports, Hengelosestraat 500, 7521 AN Enschede, Netherlands\\\email{\{l.bransen, j.vanhaaren\}@scisports.com}}

\maketitle

\begin{abstract}
Several performance metrics for quantifying the in-game performances of individual football players have been proposed in recent years. Although the majority of the on-the-ball actions during games constitutes of passes, many of the currently available metrics focus on measuring the quality of shots only. To help bridge this gap, we propose a novel approach to measure players' on-the-ball contributions from passes during games. Our proposed approach measures the expected impact of each pass on the scoreline.

\end{abstract}

\section{Introduction}

The performances of individual football players in games are hard to quantify due to the low-scoring nature of football. During major tournaments like the FIFA World Cup, the organizers\footnote{\url{https://www.fifa.com/worldcup/statistics/}} and mainstream sports media report basic statistics like distance covered, number of assists, number of saves, number of goal attempts, and number of completed passes~\cite{burnton2018best}. While these statistics provide some insights into the performances of individual football players, they largely fail to account for the circumstances under which the actions were performed. For example, successfully completing a forward pass deep into the half of the opponent is both more difficult and more valuable than performing a backward pass on the own half without any pressure from the opponent whatsoever.

In recent years, football analytics researchers and enthusiasts alike have proposed several performance metrics for individual players. Although the majority of these metrics focuses on measuring the quality of shots, there has been an increasing interest in quantifying other types of individual player actions~\cite{decroos2018hattrics,gyarmati2016qpass,power2017not}. This recent focus shift has been fueled by the availability of more extensive data and the observation that shots only constitute a small portion of the on-the-ball actions that football players perform during games~\cite{power2017not}. In the present work, we use a dataset comprising over twelve million on-the-ball actions of which only 2\% are shots. Instead, the majority of the on-the-ball actions are passes (75\%), dribbles (13\%), and set pieces (10\%).

In this paper, we introduce a novel approach to measure football players' on-the-ball contributions from passes during games. Our approach measures the expected impact of each pass on the scoreline. We value a pass by computing the difference between the expected reward of the possession sequence constituting the pass both before and after the pass. That is, a pass receives a positive value if the expected reward of the possession sequence after the pass is higher than the expected reward before the pass. Our approach employs a k-nearest-neighbor search with dynamic time warping (DTW) as a distance function to determine the expected reward of a possession sequence. Our empirical evaluation on an extensive real-world dataset shows that our approach is capable of identifying different types of impactful players like the ball-playing defender Ragnar Klavan (Liverpool FC), the advanced playmaker Mesut \"{O}zil (Arsenal), and the deep-lying playmaker Toni Kroos (Real Madrid).

\section{Dataset}
\label{section:dataset}

Our dataset comprises game data for 9,061 games in the English Premier League, Spanish LaLiga, German 1. Bundesliga, Italian Serie A, French Ligue Un, Belgian Pro League and Dutch Eredivisie. The dataset spans the 2014/2015 through 2017/2018 seasons and was provided by SciSports' partner Wyscout.\footnote{\url{https://www.wyscout.com}} For each game, the dataset contains information on the players (i.e., name, date of birth and position) and the teams (i.e., starting lineup and substitutions) as well as play-by-play event data describing the most notable events that happened on the pitch. For each event, the dataset provides the following information: timestamp (i.e., half and time), team and player performing the event, type (e.g., pass or shot) and subtype (e.g., cross or high pass), and start and end location. Table~\ref{table:data-example} shows an excerpt from our dataset showing five consecutive passes.

\begin{table}[htp!]
\centering
\caption{An excerpt from our dataset showing five consecutive passes.}
\label{table:data-example}
\begin{tabular}{rrrrrrrrrr}
\toprule
\textbf{half} & \textbf{time (s)} & \textbf{team} & \textbf{player} & \textbf{type} & \textbf{subtype} & \textbf{start\_x} & \textbf{end\_x} & \textbf{start\_y} & \textbf{end\_y} \tabularnewline
\midrule
1 & 8.642 & 679 & 217031 & 8 & 85 & 58 & 66 & 34 & 9 \tabularnewline
1 & 10.167 & 679 & 86307 & 8 & 85 & 66 & 85 & 9 & 17 \tabularnewline
1 & 11.987 & 679 & 3443 & 8 & 85 & 85 & 90 & 17 & 25 \tabularnewline
1 & 13.681 & 679 & 4488 & 8 & 80 & 90 & 89 & 25 & 44 \tabularnewline
1 & 14.488 & 679 & 3682 & 8 & 85 & 89 & 81 & 44 & 39 \tabularnewline
\bottomrule
\end{tabular}
\end{table}

\section{Approach}
\label{section:approach}

Valuing individual actions such as passes is challenging due to the low-scoring nature of football. Since football players get only a few occasions during games to earn reward from their passes (i.e., each time a goal is scored), we resort to computing the passes' expected rewards instead of distributing the actual rewards from goals across the preceding passes. More specifically, we compute the number of goals expected to arise from a given pass if that pass were repeated many times. To this end, we propose a four-step approach to measure football players' expected contributions from their passes during football games. In the remainder of this section, we discuss each step in turn.

\subsection{Constructing possession sequences}

We split the event stream for each game into a set of possession sequences, which are sequences of events where the same team remains in possession of the ball. The first possession sequence in each half starts with the kick-off. The end of one possession sequence and thus also the start of the following possession sequence is marked by one the following events: a team losing possession (e.g., due to an inaccurate pass), a team scoring a goal, a team committing a foul (e.g., an offside pass), or the ball going out of play.

\subsection{Labeling possession sequences}

We label each possession sequence by computing its expected reward. When a possession sequence \textit{does not} result in a shot, the sequence receives a value of zero. When a possession sequence \textit{does} result in a shot, the sequence receives the expected-goals value of the shot. This value reflects how often the shot can be expected to yield a goal if the shot were repeated many times. For example, a shot having an expected-goals value of 0.13 is expected to translate into 13 goals if the shot were repeated 100 times.

Building on earlier work, we train an expected-goals model to value shots~\cite{decroos2017predicting}. We represent each shot by its location on the pitch (i.e., x and y location), its distance to the center of the goal, and the angle between its location and the goal posts. We label the shots resulting in a goal as positive examples and all other shots as negative examples. We train a binary classification model that assigns a probability of scoring to each shot.

\subsection{Valuing passes}
\label{subsection:valuing-passes}

\begin{figure}[ht]
 \begin{center}
  \includegraphics[width=0.8\textwidth]{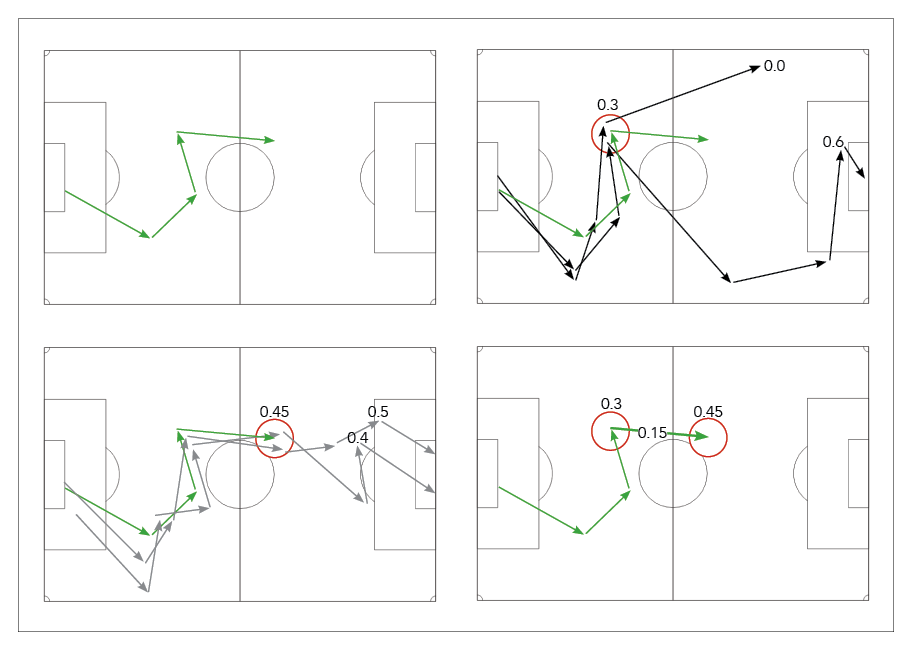}
  \caption{Visualization of our approach to value passes, where we value the last pass of the possession sequence shown in green. We obtain a pass value of 0.15, which is the difference between the value of the possession subsequence after the pass (0.45) and the value of the possession subsequence before the pass (0.30).}
  \label{fig:method}
 \end{center}
 \vspace{-20pt}
\end{figure}

We split each possession sequence into a set of possession subsequences. Each subsequence starts with the same event as the original possession sequence and ends after one of the passes in that sequence. For example, a possession sequence consisting of a pass 1, a pass 2, a dribble and a pass 3 collapses into a set of three possession subsequences. The first subsequence consists of pass 1, the second subsequence consists of pass 1 and pass 2, and the third subsequence consists of pass 1, pass 2, the dribble, and pass 3.

We value a given pass by computing the difference between the expected reward of the possession subsequence after that pass and the expected reward of the possession subsequence before that pass. Hence, the value of the pass reflects an increase or decrease in expected reward. We assume that a team can only earn reward whenever it is in possession of the ball. If the pass is the first in its possession subsequence, we set the expected reward of the possession subsequence before the pass to zero. If the pass is unsuccessful and thus marks the end of its possession subsequence, we set the expected reward of the possession subsequence after the pass to zero.

We compute the expected reward of a possession subsequence by first performing a k-nearest-neighbors search and then averaging the labels of the k most-similar possession subsequences. We use dynamic time warping (DTW) to measure the similarity between two possession subsequences~\cite{bemdt1994using}. We interpolate the possession subsequences and obtain the x and y coordinates at fixed one-second intervals. We first apply DTW to the x coordinates and y coordinates separately and then sum the differences in both dimensions.

To speed up the k-nearest-neighbors search, we reduce the number of computations by first clustering the possession subsequences and then performing DTW within each cluster. We divide the pitch into a grid of cells, where each cell is 15 meters long and 17 meters wide. Hence, a default pitch of 105 meters long and 68 meters wide yields a 7-by-4 grid. We represent each cluster as an origin-destination pair and thus obtain 784 clusters (i.e., 28 origins $\times$ 28 destinations). We assign each possession subsequence to exactly one cluster based on its start and end location on the pitch.

Figure~\ref{fig:method} shows a visualization of our approach for valuing passes. In this example, we aim to value the last pass in the possession sequence shown in green (top-left figure). First, we compute the value of the possession subsequence before the pass (top-right figure). We compute the average of the labels of the two nearest neighbors, which are 0.0 and 0.6, and obtain a value of 0.3. Second, we compute the value of the possession subsequence after the pass (bottom-left figure). We compute the average of the labels of the two nearest neighbors, which are 0.4 and 0.5, and obtain a value of 0.45. Third, we compute the difference between the value after the pass and the value before the pass to obtain a pass value of 0.15 (bottom-right figure).

\subsection{Rating players}

We rate a player by first summing the values of his passes for a given period of time (e.g., a game, a sequence of games or a season) and then normalizing the obtained sum per 90 minutes of play. We consider all types of passes, including open-play passes, goal kicks, corner kicks, and free kicks.

\section{Experimental evaluation}
\label{section:results}

In this section, we present an experimental evaluation of our proposed approach. We introduce the datasets, present the methodology, investigate the impact of the parameters, and present results for the 2017/2018 season.

\subsection{Datasets}

We split the available data presented in Section~\ref{section:dataset} into three datasets: a train set, a validation set, and a test set. We respect the chronological order of the games. Our train set covers the 2014/2015 and 2015/2016 seasons, our validation set covers the 2016/2017 season, and our test set covers the 2017/2018 season. Table~\ref{tbl:datasets} shows the characteristics of our three datasets.

\begin{table}
  \centering
  \caption{The characteristics of our three datasets.}
  \label{tbl:datasets}
  \begin{tabular}{lrrr}
    \toprule
    & \textbf{Train set} & \textbf{Validation set} & \textbf{Test set} \tabularnewline
    \midrule
    Games & {4,253} & {2,404} & {2,404} \tabularnewline
    Possession sequences & {1,878,593} & {972,526} & {970,303} \tabularnewline
    Passes & {3,425,285} & {1,998,533} & {2,023,730} \tabularnewline
    Shots & {95,381} & {53,617} & {54,311} \tabularnewline
    Goals & {9,853} & {5,868} & {5,762} \tabularnewline
    \bottomrule
  \end{tabular}
\end{table}

\subsection{Methodology}
\label{subsection:methodology}

We use the \texttt{XGBoost} algorithm to train the expected-goals model.\footnote{\url{https://xgboost.readthedocs.io/en/latest/}} After optimizing the parameters using a grid search, we set the number of estimators to 500, the learning rate to 0.01, and the maximum tree depth to 5. We use the dynamic time warping implementation provided by the \texttt{dtaidistance} library to compute the distances between the possession subsequences.\footnote{\url{https://github.com/wannesm/dtaidistance}} We do not restrict the warping paths in the distance computations.

Inspired by the work from Liu and Schulte on evaluating player performances in ice hockey, we evaluate our approach by predicting the outcomes of future games as we expect our pass values to be predictors of future performances~\cite{liu2018deepreinforcementlearning}. We predict the outcomes for 1,172 games in the English Premier League, Spanish LaLiga, German 1. Bundesliga, Italian Serie A and French Ligue Un. We only consider games involving teams for which player ratings are available for at least one player in each line (i.e., goalkeeper, defender, midfielder or striker).

We assume that the number of goals scored by each team in each game is Poisson distributed~\cite{maher1982modelling}. We use the player ratings obtained on the validation set to determine the means of the Poisson random variables representing the expected number of goals scored by the teams in the games in the test set. We compute the Poisson means by summing the ratings for the players in the starting line-up. For players who played at least 900 minutes in the 2016/2017 season, we consider their actual contributions. For the remaining players, we use the average contribution of the team's players in the same line. Since the average reward gained from passes (i.e., 0.07 goals per team per game) only reflects around 5\% of the average reward gained during games (i.e., 1.42 goals per team per game), we transform the distribution over the total player ratings per team per game to follow a similar distribution as the average number of goals scored by each team in each game in the validation set. We compute the probabilities for a home win, a draw, and an away win using the Skellam distribution~\cite{karlis2008bayesian}.

\subsection{Impact of the parameters}

We now investigate how the clustering step impacts the results and what the optimal number of neighbors in the k-nearest-neighbors search is.

\subsubsection{Impact of the clustering step.}

For an increasing number of possession sequences, performing the k-nearest-neighbors search quickly becomes prohibitively expensive. For example, obtaining results on our test set would require over 1.8 trillion distance computations (i.e., 1,878,593 possession sequences in the train set $\times$ 970,303 possession sequences in the test set). To reduce the number of distance computations, we exploit the observation that possession sequences starting or ending in entirely different locations on the pitch are unlikely to be similar. For example, a possession sequence starting in a team's penalty area is unlikely to be similar to a possession sequence starting in the opponent's penalty area. More specifically, as explained in Section~\ref{subsection:valuing-passes}, we first cluster the possession sequences according to their start and end locations and then perform the k-nearest-neighbors search within each cluster.

To evaluate the impact of the clustering step on our results, we arbitrarily sample 100 games from the train set and 50 games from the validation set. The resulting train and validation subsets consist of 68,907 sequences and 35,291 sequences, respectively. Table~\ref{tbl:clustering} reports the total runtimes, the number of clusters, and the average cluster size for three settings: no clustering, clustering with grid cells of 15 by 17 meters, and clustering with grid
cells of 5 by 4 meters. As expected, clustering the possession sequences speeds up our approach considerably.

\begin{figure}[H]
	\vspace{-10pt}
	\begin{center}
		\includegraphics[width=0.7\textwidth]{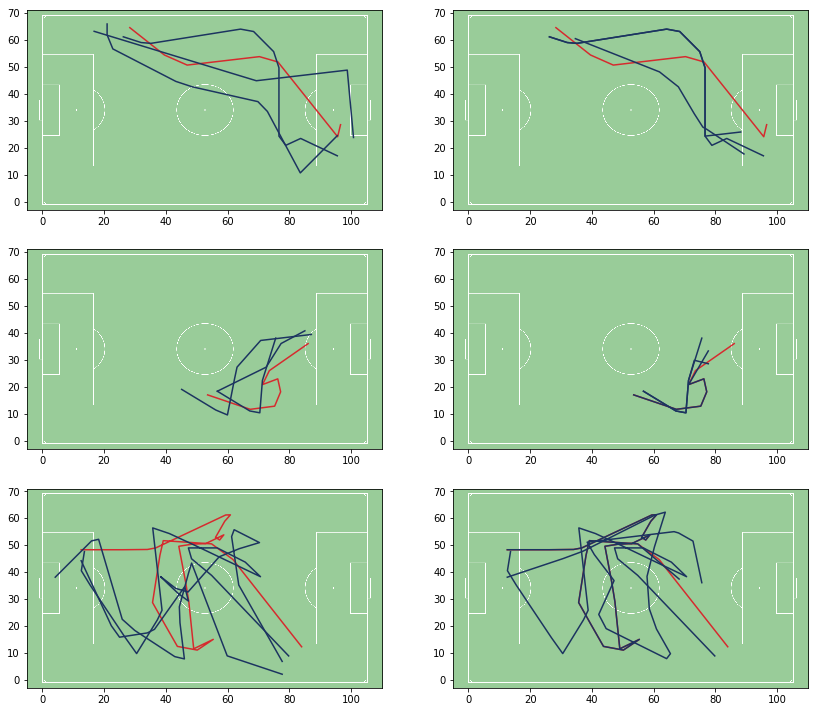}
		\caption{Visualization of the three nearest neighbors obtained for three randomly sampled possession sequences in both the clustering setting with grid cells of 15 by 17 meters and the no-clustering setting. The results for the former setting are on the left, whereas the results for the latter setting are on the right.}
		\label{fig:sequences_preclus}
	\end{center}

\end{figure}

In addition, we also investigate the impact of the clustering step in a more qualitative fashion. We randomly sample three possession sequences and 100 games comprising 32,245 possession sequences from our training set. We perform a three-nearest-neighbors search in both the no-clustering setting and the clustering setting with grid cells of 15 by 17 meters. Figure~\ref{fig:sequences_preclus} shows the three nearest neighbors for each of the three possession sequences in both settings, where the results for the clustering setting are shown on the left and the results for the no-clustering setting are shown on the right. Although the obtained possession sequences are different, the three-nearest-neighbors search obtains highly similar neighbors in both settings.

\begin{table}[H]
	\centering
	\caption{The characteristics of three settings: no clustering, clustering with grid cells of 15 by 17 meters, and clustering with grid cells of 5 by 4 meters.}
	\label{tbl:clustering}
	\begin{tabular}{lrrr}
		\toprule
		& \textbf{No clustering} & \textbf{Cell: 15$\times$17} & \textbf{Cell: 5$\times$4}\tabularnewline
		\midrule
		Total runtime & {270 minutes} & {12 minutes} & {150 minutes} \tabularnewline
		Number of clusters & 1 & 784 & 127,449 \tabularnewline
		Average cluster size & {68,907} & {87.89} & {0.54} \tabularnewline
		\bottomrule
	\end{tabular}
\end{table}

\subsubsection{Optimal number of neighbors in the k-nearest-neighbors search.}

We investigate the optimal number of neighbors in the k-nearest-neighbors search to value passes. We try the following values for the parameter $k$: 1, 2, 5, 10, 20, 50 and 100. We predict the outcomes of the games in the test set as explained in Section~\ref{subsection:methodology}. Table~\ref{tbl:loglosses} shows the logarithmic losses for each of the values for $k$. We find that 10 is the optimal number of neighbors.

In addition, we compare our approach to two baseline approaches. The first baseline approach is the pass accuracy. The second baseline approach is the prior distribution over the possible game outcomes, where we assign a probability of 48.42\% to a home win, 28.16\% to an away win, and 23.42\% to a draw. Our approach outperforms both baseline approaches.

\subsection{Results}

We now present the players who provided the highest contributions from passes during the 2017/2018 season. We present the overall ranking as well as the top-ranked players under the age of 21. Furthermore, we investigate the relationship between a player's average value per pass and his total number of passes per 90 minutes as well as the distribution of the player ratings per position.

\begin{table}[H]
	\centering
	\caption{Logarithmic losses for predicting the games in the test set for different numbers of nearest neighbors $k$ in order of increasing logarithmic loss.}
	\label{tbl:loglosses}
	\begin{tabular}{lr}
		\toprule
		\textbf{Setting} & \textbf{Logarithmic loss} \tabularnewline
		\midrule
		$k=10$ & {1.0521} \tabularnewline
		$k=5$ & {1.0521} \tabularnewline
		$k=20$ & {1.0528} \tabularnewline
		$k=2$ & {1.0560} \tabularnewline
		$k=50$ & {1.0579} \tabularnewline
		$k=100$ & {1.0594} \tabularnewline
		$k=1$ & {1.0725} \tabularnewline
		Pass accuracy & {1.0800} \tabularnewline
		Prior distribution & {1.0860}
		\tabularnewline
		\bottomrule
	\end{tabular}
\end{table}

Following the experiments above, we set the number of nearest neighbors $k$ to 10 and perform clustering with grid cells of 15 meters by 17 meters. We compute the expected reward per 90 minutes for the players in the 2017/2018 season (i.e., the test set) and perform the k-nearest-neighbors search to value their passes on all other seasons (i.e., the train and validation set).

Table~\ref{tbl:ranking-1718} shows the top-ten-ranked players who played at least 900 minutes during the 2017/2018 season in the English Premier League, Spanish LaLiga, German 1. Bundesliga, Italian Serie A, and French Ligue Un. Ragnar Klavan, who is a ball-playing defender for Liverpool FC, tops our ranking with an expected contribution per 90 minutes of 0.1133. Furthermore, Arsenal's advanced playmaker Mesut \"{O}zil ranks second, whereas Real Madrid's deep-lying playmaker Toni Kroos ranks third.

Table~\ref{tbl:ranking-1718-talents} shows the top-five-ranked players under the age of 21 who played at least 900 minutes during the 2017/2018 season in Europe's top-five leagues, the Dutch Eredivisie or the Belgian Pro League. Teun Koopmeiners (AZ Alkmaar) tops our ranking with an expected contribution per 90 minutes of 0.0806. Furthermore, Real Madrid-loanee Martin {\O}degaard (SC Heerenveen) ranks second, whereas Nikola Milenkovi\'{c} (ACF Fiorentina) ranks third.

\begin{table}[H]
	\vspace{-10pt}
	\centering
	\caption{The top-ten-ranked players who played at least 900 minutes during the 2017/2018 season in Europe's top-five leagues.}
	\label{tbl:ranking-1718}
	\begin{tabular}{rllr}
		\toprule
		\textbf{Rank} & \textbf{Player} & \textbf{Team} & \textbf{Contribution P90} \tabularnewline
		\midrule
		1 & Ragnar Klavan & Liverpool FC & 0.1133 \tabularnewline
		2 & Mesut \"{O}zil & Arsenal & 0.1034 \tabularnewline
		3 & Toni Kroos & Real Madrid & 0.0943  \tabularnewline
		4 & Manuel Lanzini & West Ham United & 0.0892 \tabularnewline
		5 & Joan Jord\'{a}n & SD Eibar & 0.0830 \tabularnewline
		6 & Esteban Granero & Espanyol & 0.0797 \tabularnewline
		7 & Nuri Sahin & Borussia Dortmund & 0.0796 \tabularnewline
		8 & Mahmoud Dahoud & Borussia Dortmund & 0.0775 \tabularnewline
		9 & Granit Xhaka & Arsenal & 0.0774 \tabularnewline
		10 & Faouzi Ghoulam & SSC Napoli & 0.0765 \tabularnewline
		\bottomrule
	\end{tabular}
	\vspace{-10pt}
\end{table}

\begin{table}[H]
	\vspace{-10pt}
  \centering
  \caption{The top-five-ranked players under the age of 21 who played at least 900 minutes during the 2017/2018 season in Europe's top-five leagues, the Dutch Eredivisie or the Belgian Pro League.}
  \label{tbl:ranking-1718-talents}
  \begin{tabular}{rllr}
    \toprule
    \textbf{Rank} & \textbf{Player} & \textbf{Team} & \textbf{Contribution P90} \tabularnewline
    \midrule
    1 & Teun Koopmeiners & AZ Alkmaar & 0.0806 \tabularnewline
    2 & Martin {\O}degaard & SC Heerenveen & 0.0639 \tabularnewline
    3 & Nikola Milenkovi\'{c} & ACF Fiorentina & 0.0617  \tabularnewline
    4 & Sander Berge & KRC Genk & 0.0601 \tabularnewline
    5 & Maximilian W\"{o}ber & Ajax & 0.0599 \tabularnewline
    \bottomrule
  \end{tabular}
	\vspace{-10pt}
\end{table}

\begin{figure}[htb]
  \begin{center}
    \includegraphics[width=0.8\textwidth]{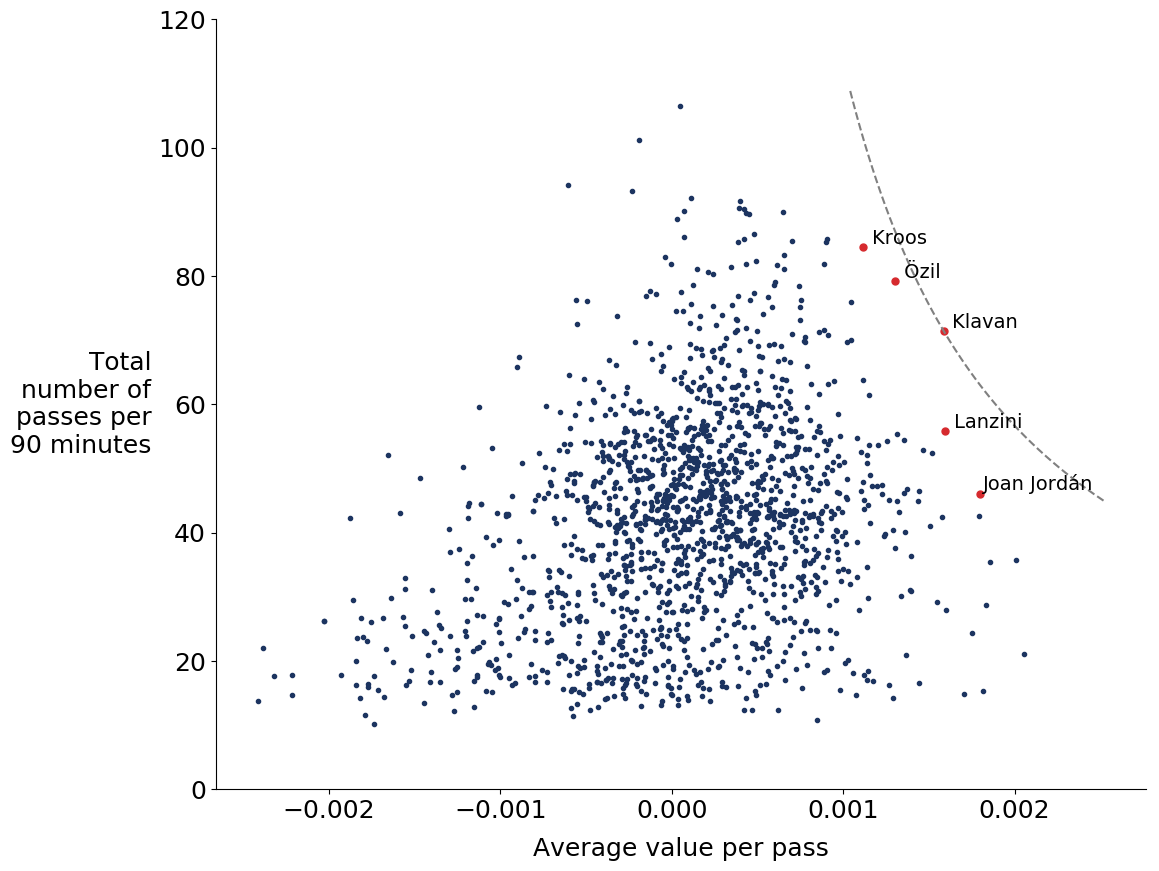}
    \caption{Scatter plot showing the correlation between a player's average value per pass and his total number of passes per 90 minutes. The five players with the highest contribution per 90 minutes are highlighted in red.}
    \label{fig:top10scatter}
  \end{center}
\end{figure}

Figure~\ref{fig:top10scatter} shows whether players earn their pass contribution by performing many passes per 90 minutes or by performing high-value passes. The five players with the highest contribution per 90 minutes are highlighted in red. While Lanzini and Joan Jord\'{a}n do not perform many passes per 90 minutes, they obtain a rather high average value per pass. The dotted line drawn through Klavan contains all points with the same contribution per 90 minutes as him.

\begin{figure}[htb]
  \centering
  \subfloat{\includegraphics[width=.4\textwidth]{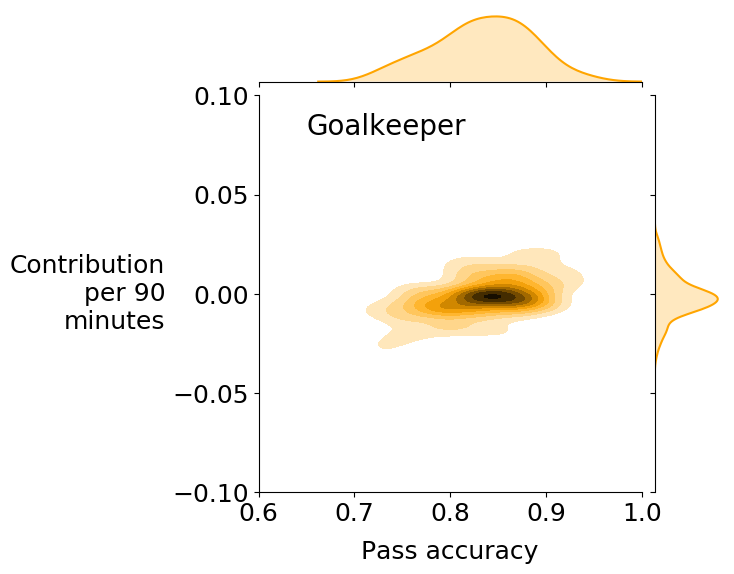}}\quad
  \subfloat{\includegraphics[width=.4\textwidth]{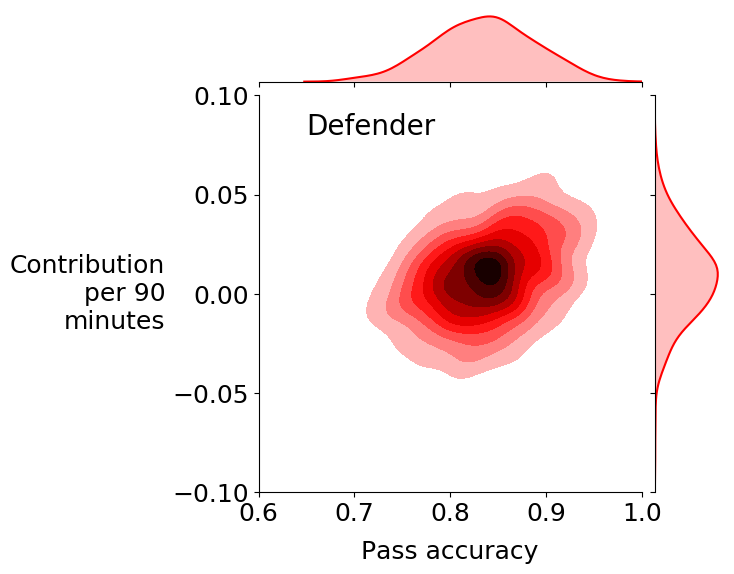}}\\
  \subfloat{\includegraphics[width=.4\textwidth]{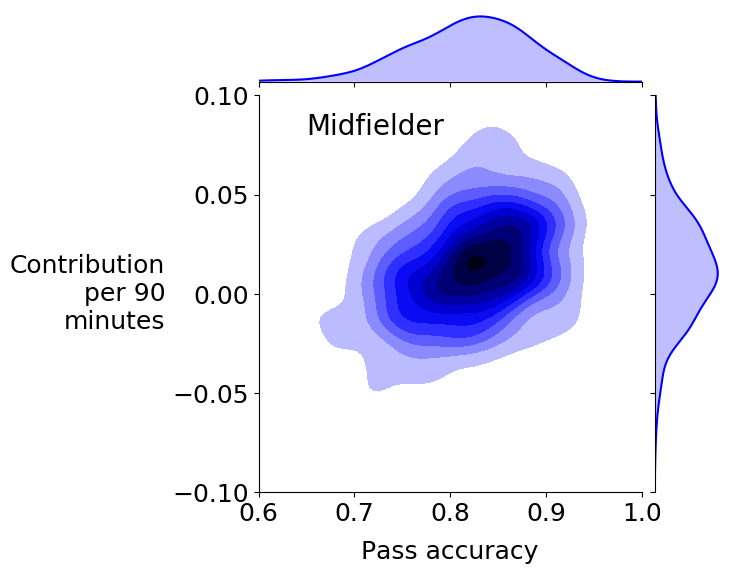}}\quad
  \subfloat{\includegraphics[width=.4\textwidth]{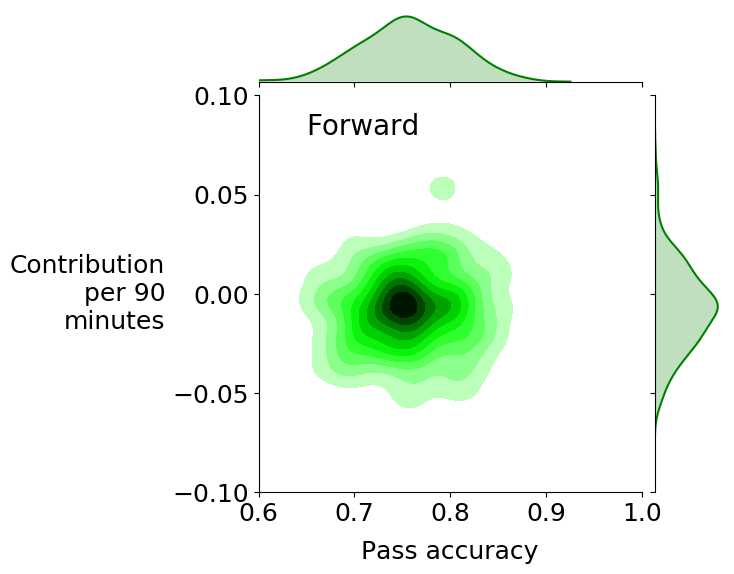}}
  \caption{Density plots per position showing the correlation between the pass accuracy and the pass contribution per 90 minutes.}
  \label{fig:positions_kde}
  \vspace{-10pt}
\end{figure}

Figure~\ref{fig:positions_kde} presents a comparison between a player's pass accuracy and pass contribution per 90 minutes. In terms of pass accuracy, forwards rate low as they typically perform passes in more crowded areas of the pitch, while goalkeepers rate high. In terms of pass contribution, goalkeepers rate low, while especially midfielders rate high.

\subsection{Application: Replacing Manuel Lanzini}

We use our pass values to find a suitable replacement for Manuel Lanzini. The Argentine midfielder, who excelled at West Ham United throughout the 2017/2018 season, ruptured his right knee's anterior cruciate ligament while preparing for the 2018 FIFA World Cup. West Ham United are expected to sign a replacement for Lanzini, who will likely miss the entire 2018/2019 season.

To address this task, we define a ``Lanzini similarity function'' that accounts for a player's pass contribution per 90 minutes, number of passes per 90 minutes and pass accuracy. We normalize each of the three pass metrics before feeding them into the similarity function. Manuel Lanzini achieves a high pass contribution per 90 minutes despite a low pass accuracy, which suggests that the midfielder prefers high-risk, high-value passes over low-risk, low-value passes.

Table~\ref{tbl:lanzini} shows the five most-similar players to Lanzini born after July 1st, 1993, who played at least 900 minutes in the 2017/2018 season. Mahmoud Dahoud (Borussia Dortmund) tops the list ahead of Joan Jord\'{a}n (SD Eibar) and Naby Ke\"{i}ta (RB Leipzig), who moved to Liverpool during the summer of 2018.

\begin{table}[H]
	\vspace{-10pt}
  \centering
  \caption{The five most-similar players to Manuel Lanzini born after July 1st, 1993, who played at least 900 minutes in the 2017/2018 season.}
  \label{tbl:lanzini}
  \begin{tabular}{rllr}
    \toprule
    \textbf{Rank} & \textbf{Player} & \textbf{Team} & \textbf{Similarity score} \tabularnewline
    \midrule
    1 & Mahmoud Dahoud & Borussia Dortmund & 0.9955 \tabularnewline
    2 & Joan Jord\'{a}n & SD Eibar & 0.9881 \tabularnewline
    3 & Naby Ke\"{i}ta & RB Leipzig & 0.9794  \tabularnewline
    4 & Dominik Kohr & Bayer 04 Leverkusen& 0.9717 \tabularnewline
    5 & Medr\'{a}n & Deportivo Alav\'{e}s & 0.9591 \tabularnewline
    \bottomrule
  \end{tabular}
\end{table}

\section{Related work}

Although the focus of the football analytics community has been mostly on developing metrics for measuring chance quality, there has also been some work on quantifying other types of actions like passes. Power et al.~\cite{power2017not} objectively measure the expected risk and reward of each pass using spatio-temporal tracking data. Gyarmati and Stanojevic~\cite{gyarmati2016qpass} value each pass by quantifying the value of having the ball in the origin and destination location of the pass using event data. Although their approach is similar in spirit to ours, our proposed approach takes more contextual information into account to value the passes.

Furthermore, there has also been some work in the sports analytics community on more general approaches that aim to quantify several different types of actions~\cite{cervone2014pointwise,decroos2018hattrics,decroos2017starss,liu2018deepreinforcementlearning,schulte2015qfunction}. Decroos et al.~\cite{decroos2017starss} compute the value of each on-the-ball action in football (e.g., a pass or a dribble) by computing the difference between the values of the post-action and pre-action game states. Their approach distributes the expected reward of a possession sequence across the constituting actions, whereas our approach computes the expected reward for each pass individually. Cervone et al.~\cite{cervone2014pointwise} propose a method to predict points and to value decisions in basketball. Liu and Schulte as well as Schulte et al.~\cite{liu2018deepreinforcementlearning,schulte2015qfunction} present a deep reinforcement learning approach to address a similar task for ice hockey.

\section{Conclusion}

This paper introduced a novel approach for measuring football players' on-the-ball contributions from passes using play-by-play event data collected during football games. Viewing a football game as a series of possession sequences, our approach values each pass by computing the difference between the values of its constituting possession sequence before and after the pass. To value a possession sequence, our approach combines a k-nearest-neighbor search with dynamic time warping, where the value of the possession sequence reflects its likeliness of yielding a goal.

In the future, we aim to improve our approach by accounting for the strength of the opponent and more accurately valuing the possession sequences by taking more contextual information into account. To compute the similarities between the possession sequences, we also plan to investigate spatio-temporal convolution kernels (e.g.,~\cite{knauf2014spatio}) as an alternative for dynamic time warping  and to explore more sophisticated techniques for clustering the possession sequences.

\clearpage

\bibliographystyle{splncs04}
\bibliography{paper}

\end{document}